\documentclass[twocolumn,showpacs,preprintnumbers,amsmath,amssymb]{revtex4}

\usepackage{graphicx}
\usepackage{dcolumn}
\usepackage{bm}

\begin{document}

\preprint{}
\input{epsf.tex}

\epsfverbosetrue

\title{Coupling of polaritons to vibrational modes of ultracold atoms in an optical lattice}

\author{Hashem Zoubi, and Helmut Ritsch}

\affiliation{Institut f\"{u}r Theoretische Physik, Universit\"{a}t Innsbruck, Technikerstrasse 25, A-6020 Innsbruck, Austria}


\begin{abstract}
The coupling of internal electronic excitations to vibrational modes of the external motion of ultracold atoms in an optical lattice is studied here in using a perturbation expansion in small atomic displacements. In the Mott insulator case with one atom per site, the resonance dipole-dipole coupling between neighboring sites can induce emission and absorption of vibrational quanta. Within a cavity in the strong exciton-photon coupling regime such coupling results in polariton-vibration interactions, which create a significant thermalization mechanism for polaritons toward their minimum energy, and leading to motional heating of the lattice atoms.
\end{abstract}





\maketitle

\section{Introduction}

The superfluid to the Mott insulator quantum phase transition in a system of ultracold boson atoms in an optical lattice has been first predicted theoretically \cite{Jaksch}, and then realized experimentally \cite{Bloch}. The optical lattice is formed off standing waves of counter propagating laser beams with lattice constant of half laser wave length. The ultracold atoms are loaded into such optical lattice, and in varying the laser intensity the superfluid-Mott insulator phase transition can be achieved, and in the Mott insulator phase a fixed number of atoms per site is obtained, where the transition is well described by the Bose-Hubbard model \cite{Jaksch}. For deep optical lattice potentials the on-site lowest motional states can be described by harmonic oscillators, and the transitions of the atoms between these vibrational states are usually induced by Raman scattering of light or via scattering interactions during atom hopping among different sites \cite{Zoller}. We study here a different mechanism for the excitation and de-excitation of vibrational modes which is induced by resonance dipole-dipole interactions. The discussion is limited to two-level atoms in a deep optical lattice prepared in the Mott insulator phase with one atom per site, and we assume ground and excited state optical lattice potentials with minima at the same positions but with different depth.

We already studied collective electronic excitations (excitons \cite{Zoubi}) in such a system for the cases of one and two atoms per site. For an optical lattice, which is located between two cavity mirrors \cite{Maschler}, we showed that in the strong coupling regime, excitons and photons are coherently mixed to form cavity polaritons \cite{HashemA,HashemB,HashemC}. In these works we considered only the case of atoms localized in the lowest vibrational state, i.e. the first Bloch band. However, in more realistic optical lattices, dynamical excitation of atoms to higher vibrational states will occur. In the present work we aim to include higher vibrational states which are induced through their coupling to excitons. As optical lattices in the Mott insulator phase are similar to molecular crystals, the formalism adopted here is in the spirit of the coupling of Frenkel excitons and optical phonons in molecular crystals \cite{Davydov,Agranovich}, but some caution is needed to identify the correct physical analogies between optical phonons in molecular crystals and the vibrational states in optical lattices. In our case the interaction is mediated by resonance dipole-dipole coupling and is treated in the perturbation theory. Furthermore, polariton-vibration interactions emerge through the polariton excitonic part, and serve as a significant source of polaritons relaxation toward their minimum energy, and as an important mechanism for excitation and de-excitation of vibrational modes.

The paper is organized as follows. In section 2 we present electronic excitations and vibrational modes of cold atoms excited to higher Bloch bands in an optical lattice. And in section 3 we derive the coupling of electronic excitations to vibrational modes. We derive the cavity-polaritons coupling to the vibrational modes in section 4. A summary appears in section 5.

\section{Electronic excitations and vibrational states for an optical lattice}

The electronic excitations for ultracold atoms of an optical lattice in the Mott insulator phase can be represented by the Hamiltonian
\begin{equation}
H=\sum_i\hbar\omega_A\ B_i^{\dagger}B_i+\sum_{i,j}\hbar J_{ij}\ B_i^{\dagger}B_j,
\end{equation}
where $B_i^{\dagger},\ B_i$ are the creation and annihilation operators of an excitation at lattice position $i$. In the case of a single electronic excitation, the operators can be taken to obey the Bose commutation relation $[B_i,B_j^{\dagger}]=\delta_{ij}$. The internal atomic transition frequency is $\omega_A$. An internal electronic excitation can be exchanged among atoms at different sites $i$ and $j$, which is induced by dipole-dipole interactions and can be parametrized by a coupling integral $J_{ij}$. It is now essential to note that this coupling depends on the atomic displacements ${\bf u}^{\lambda m}_i$ and thus on the local vibrational excitation. Here ${\bf u}^{\lambda m}_i$ is the average change of the atomic size at site $i$, of position ${\bf n}_i$, in the $\lambda$ internal state, for the ground and excited state energies $(\lambda=e,g)$, due to the excitation of the atom to a higher vibrational state with $m$ quanta, and where ${\bf u}^{\lambda m}_i$ is measured relative to ${\bf u}^{\lambda 0}_i$ of the vibration-less state. The dependence of the on-site transition energy $\hbar\omega_A$ on the vibrational state is neglected. As long as ${\bf u}^{\lambda m}_i$ is a small distance relative to the lattice constant $a$, which is the case for the lowest vibrational modes, to lowest orders in this small perturbation we can split the Hamiltonian in the form $H=H_{ex}+H_{vib}+H_{ex-vib}$. Here $H_{ex}$ is the internal excitation Hamiltonian, which is obtained for atoms in the ground vibrational states, $H_{vib}$ is the atom vibration Hamiltonian, for excited and ground state atoms, and the coupling Hamiltonian $H_{ex-vib}$ is derived perturbatively for atoms excited to higher vibrational states.

To zeroth order, for atoms in the ground vibrational states, we thus get
\begin{equation}
H_{ex}=\sum_i\hbar\omega_A\ B_i^{\dagger}B_i+\sum_{i,j}\hbar J^0_{ij}\ B_i^{\dagger}B_j,
\end{equation}
where $J^0_{ij}$ is the transfer parameter among atoms in the lowest vibrational states.

The Hamiltonian can be diagonalized by using the transformation into the quasi-momentum space
\begin{equation} \label{TRANS}
B_i=\frac{1}{\sqrt{N}}\sum_{\bf k}e^{i{\bf k}\cdot{\bf n}_i}B_{\bf k},
\end{equation}
where $N$ is the number of sites, and ${\bf k}$ is the in-plane wave vector for 2D optical lattice. We obtain the exciton Hamiltonian \cite{HashemA}
\begin{equation} \label{EXXX}
H_{ex}=\sum_{\bf k}\hbar\omega_a({\bf k})\ B_{\bf k}^{\dagger}B_{\bf k}, 
\end{equation}
where the exciton dispersion is $\omega_a({\bf k})=\omega_A+\sum_{\bf L}J({\bf L})e^{i{\bf k}\cdot{\bf L}}$, with ${\bf L}={\bf n}_i-{\bf n}_j$ and $J({\bf L})=J^0_{ij}$. The condition for the energy transfer among lattice sites, and hence the formation of collective electronic excitations (excitons), is to have a dipole-dipole coupling larger than the excited state line width $\Gamma$, that is $J>\Gamma$, (more details in \cite{HashemA}).

Vibrational excitations of atoms at the ground and excited states are described by the two harmonic oscillators Hamiltonian
\begin{equation}
H_{vib}=\sum_i\hbar\omega_v^g\ b_i^{\dagger}b_i+\sum_i\hbar\omega_v^e\ c_i^{\dagger}c_i,
\end{equation}
where $\omega_v^g$ and $\omega_v^e$ are the vibration frequency for ground and excited state atoms, respectively. $b_i^{\dagger},\ b_i$ and $c_i^{\dagger},\ c_i$ are the creation and annihilation operators of a vibration mode at site $i$ for ground and excited state atoms, respectively. The atomic displacement operators are
\begin{equation}
\hat{x}_i^g=\sqrt{\frac{\hbar}{2m\omega_v^g}}\left(b_i+b_i^{\dagger}\right)\ ,\ \hat{x}_i^e=\sqrt{\frac{\hbar}{2m\omega_v^e}}\left(c_i+c_i^{\dagger}\right),
\end{equation}
where $m$ is the atomic mass.

Here, ground and excited state atoms are considered as two different kinds of bosons \cite{Chen}, where each has its own optical lattice potential. In the transition between the ground and the excited state one kind of bosons is destroyed and another created. Hence we use two independent harmonic oscillators for the ground and excited vibrational excitations. Note that if we have physical allowed initial states, then the system dynamics due to the following interaction Hamiltonian, derived in the perturbation theory, will not couple to any un-physical states present in the Hilbert space.

\section{Interactions of electronic excitations to vibrational states}

To first order in the perturbation series with respect to ${\bf u}^{\lambda}_i$, we derive an approximative excitation-vibration coupling Hamiltonian $H_{ex-vib}$. Processes in first order include only single vibrational quantum, and we drop the index $(m)$. To treat processes of more than a single vibrational quantum one needs to consider higher order terms of this perturbation series. The present perturbation method \cite{Mahan} allows one to include any number of vibrational quanta, (up to the limit of the harmonic oscillator approximation).

The interaction Hamiltonian of the transfer term then reads
\begin{eqnarray}
H_{ex-vib}^{II}&=&\sum_{i,j}\hbar\left[F^{ei}_{ij}\ c_i^{\dagger}+F^{gi}_{ij}\ b_i\right. \nonumber \\
&+&\left.F^{ej}_{ij}\ c_j+F^{gj}_{ij}\ b_j^{\dagger}\right]\ B_i^{\dagger}B_j,
\end{eqnarray}
where the coupling parameter is
\begin{equation} \label{TranCoup}
F^{\lambda i}_{ij}=\sqrt{\frac{\hbar}{2m\omega_v^{\lambda}}}\left\{\frac{\partial J_{ij}}{\partial {\bf u}^{\lambda}_i}\right\}_{{\bf u}^{\lambda}_i=0},
\end{equation}
which is related to the derivative of $J_{ij}$. Furthermore, we assume the excitation-vibration coupling parameters to be site independent, by defining $F^{\lambda}_{ij}\equiv F^{\lambda i}_{ij}$. The excitation-vibration coupling Hamiltonian is written as
\begin{equation}
H_{ex-vib}=\sum_{i,j}\hbar\left[F^{e}_{ij}\left(c_i^{\dagger}+c_j\right)+F^{g}_{ij}\left(b_i+b_j^{\dagger}\right)\right]\ B_i^{\dagger}B_j.
\end{equation}
This term describes creation and destruction of vibrations, induced by excitation transfer among different sites. Two of the four processes are plotted in figures (1-2). The process in figure (1) show emission of ground state vibration at site $j$. While the process in figure (2) show absorption of excited state vibration at site $j$. In the process of figure (1), the initial states are $|g_i,0_v^g\rangle$ and $|e_j,0_v^e\rangle$ with a dipole-dipole coupling, say, $J_I$. The final states are $|e_i,0_v^e\rangle$ and $|g_j,1_v^g\rangle$ with a dipole-dipole coupling, say, $J'_I$. Such differences in the coupling parameters induce the emission and absorption of vibrational modes, as it is clear from the factor $\left\{\frac{\partial J_{ij}}{\partial {\bf u}^{\lambda}_i}\right\}_{{\bf u}^{\lambda}_i=0}$ in the coupling parameter. Similar explanations hold for the other processes. The discussion is limited here to the first order of the perturbation series, where the emission and absorption processes are between the ground and the first excited vibrational state with a single quanta. Higher orders include processes with more than a single vibrational quanta.
\begin{figure}
\centerline{\epsfxsize=8cm \epsfbox{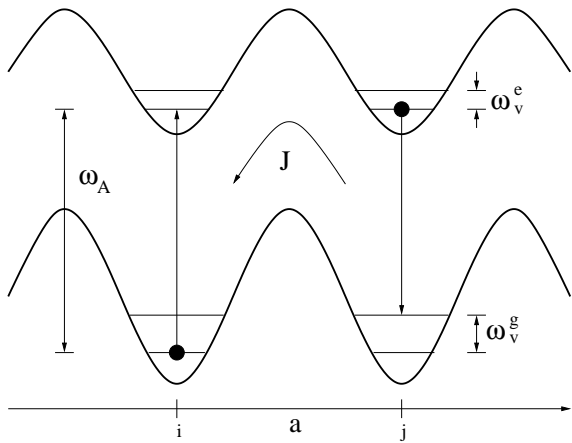}}
\caption{The term $F^{g}_{ij}\ B_i^{\dagger}B_j\ b_j^{\dagger}$, represent the process of the emission of ground state vibration at site $j$.}
\end{figure}
\begin{figure}
\centerline{\epsfxsize=8cm \epsfbox{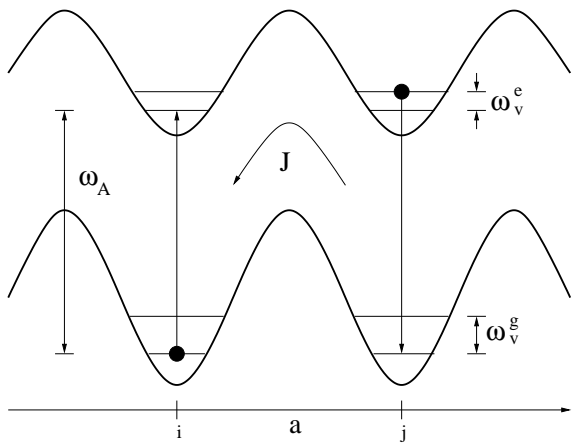}}
\caption{The term $F^{e}_{ij}\ B_i^{\dagger}B_j\ c_j$, represent the process of the absorption of ground state vibration at site $j$.}
\end{figure}

The interaction Hamiltonian can be rewritten in momentum space representation. The excitation operators, as before, casts into exciton operators in using transformation (\ref{TRANS}), with the exciton Hamiltonian of Eq.(\ref{EXXX}). The vibration operators, even though represent on-site localized vibrations, can be transformed formally into the momentum space, by applying the transformations
\begin{equation}
b_i=\frac{1}{\sqrt{N}}\sum_{\bf q}e^{i{\bf q}\cdot{\bf n}_i}b_{\bf q}\ , \ c_i=\frac{1}{\sqrt{N}}\sum_{\bf q}e^{i{\bf q}\cdot{\bf n}_i}c_{\bf q},
\end{equation}
to get
\begin{equation}
H_{vib}=\sum_{\bf q}\hbar\omega_v^g\ b_{\bf q}^{\dagger}b_{\bf q}+\sum_{\bf q}\hbar\omega_v^e\ c_{\bf q}^{\dagger}c_{\bf q},
\end{equation}
which have flat dispersions, namely $q$-independent. The exciton-vibration coupling now reads
\begin{eqnarray} \label{ExVib}
H_{ex-vib}&=&\sum_{\bf k,q}\hbar\left\{F^{e}({\bf k+q})\ c_{\bf q}+F^{g}({\bf k})\ b_{\bf q}\right. \nonumber \\
&+&\left.F^{e}({\bf k})\ c_{\bf -q}^{\dagger}+F^{g}({\bf k+q})\ b_{\bf -q}^{\dagger}\right\}\ B_{\bf k+q}^{\dagger}B_{\bf k},\nonumber \\
\end{eqnarray}
where
\begin{equation}
F^{\lambda}({\bf k})=\frac{1}{\sqrt{N}}\sum_{\bf L}F^{\lambda}({\bf L})e^{i{\bf k}\cdot{\bf L}}.
\end{equation}
The Hamiltonian $H_{ex-vib}$ describes scattering of excitons between different wave vectors by the emission and absorption of a vibration. Such a process is possible in the limit of $J\gg \omega_v^{\lambda}$, where the exciton band width is larger than the vibration energy. The scattering conserves energy and momentum, and as the vibration dispersion is a flat one, the vibration can absorb any momentum amount of the exciton.

\section{Interactions of cavity polaritons to vibrational states}

Here we derive the cavity-polaritons coupling to the vibrational modes which results of the excitation-vibration interaction. We start in presenting the cavity photons and their coupling to the excitons in an optical lattice, (more details in \cite{HashemA}). We take the optical lattice to be located in the middle between, and parallel to, two perfect cavity mirrors, as plotted in figure (3). The electromagnetic field is free in the cavity plane with in-plane wave vector ${\bf k}$, and quantized in the perpendicular direction with the discrete modes $k_z=n\pi/L$, where $n=1,2,\cdots$. We restrict our considerations to only the mode $n=1$, that is close to resonance with the atomic electronic excitation, and we concentrate in a fixed cavity photon polarization. The cavity Hamiltonian thus is given by $H_c=\sum_{\bf k}\hbar\omega_c(k)\ a^{\dagger}_{\bf k}a_{\bf k}$, where $a^{\dagger}_{\bf k},\ a_{\bf k}$ are the creation and annihilation boson operators of a cavity photon with in-plane wave vector ${\bf k}$. The corresponding cavity photon dispersion thus reads $\omega_c(k)=\frac{c}{\sqrt{\epsilon}}\sqrt{k^2+\left(\frac{\pi}{L}\right)^2}$, where $L$ is the distance between the cavity mirrors, and $\epsilon \approx 1$ is the cavity medium dielectric constant. 

\begin{figure}
\centerline{\epsfxsize=8cm \epsfbox{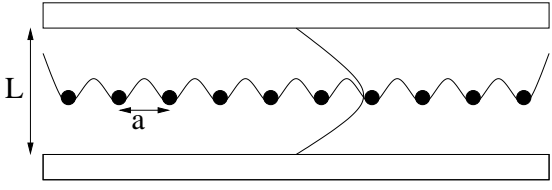}}
\caption{Schematic plot of the optical lattice between the two cavity mirrors. The cavity mode is plotted.}
\end{figure}

The coupled exciton-photon Hamiltonian is
\begin{eqnarray}
H&=&\sum_{\bf k}\hbar\left\{\omega_a\ B_{\bf k}^{\dagger}B_{\bf k}+\omega_c(k)\ a^{\dagger}_{\bf k}a_{\bf k}\right. \nonumber \\
&+&\left.f_k\ B^{\dagger}_{\bf k}a_{\bf k}+f^{\ast}_k\ a^{\dagger}_{\bf k}B_{\bf k}\right\},
\end{eqnarray}
where due to in-plane translational symmetry, the coupling is between excitons and photons with the same in-plane wave vector, and the coupling parameter is given by $\hbar f_k=-i\sqrt{\frac{\hbar\omega_{c}(k)N\mu^2}{2LS\epsilon_0}}$. As the exciton-photon coupling take place at small wave vectors, the exciton dispersion can be neglected, and we used $\omega_a(k)\approx \omega_a$. In the strong coupling regime, where the coupling is larger than both the exciton and photon line width, the excitons and photons are coherently mixed to form the new system diagonal eigenmodes which are called polaritons. The above Hamiltonian can be easily diagonalized to give
\begin{equation}
H=\sum_{{\bf k}r}\hbar\Omega_r(k)\ A^{\dagger}_{{\bf k}r}A_{{\bf k}r},
\end{equation}
which exhibits two polariton branches, with the two dispersions
\begin{equation}
\Omega_{\pm}(k)=\frac{\omega_{c}(k)+\omega_a}{2}\pm\Delta_k,
\end{equation}
where $\Delta_k=\sqrt{\delta_k^2+|f_k|^2}$, and we defined the exciton-photon detuning by $\delta_k=(\omega_{c}(k)-\omega_a)/2$. The splitting between the two polariton branches at wave vector ${\bf k}$ is $2\Delta_k$ and the splitting at the exciton-photon intersection point, where $\delta_k=0$, is $2|f_k|$ which corresponds to the vacuum Rabi splitting. The polariton operators are defined by
\begin{equation}
A_{{\bf k}\pm}=X_{k}^{\pm}B_{\bf k}+Y_{k}^{\pm}a_{{\bf k}},
\end{equation}
with the exciton and photon amplitudes
\begin{equation}
X_{k}^{\pm}=\pm\sqrt{\frac{\Delta_k\mp\delta_k}{2\Delta_k}}\ ,\ Y_{k}^{\pm}=\frac{f_{k}}{\sqrt{2\Delta_k(\Delta_k\mp\delta_k)}}.
\end{equation}
At the exciton-photon intersection point the polaritons are half exciton and half photon, that is $|X_{k}^{\pm}|^2=|Y_{k}^{\pm}|^2=1/2$. At large wave vectors the lower branch becomes excitonic, that is $|X_{k}^{-}|^2\approx1,\ |Y_{k}^{-}|^2\approx0$, and the upper branch becomes photonic, that is $|X_{k}^{+}|^2\approx0,\ |Y_{k}^{+}|^2\approx1$.

The exciton-vibration coupling yields polariton-vibration coupling. In using the inverse transformation $B_{\bf k}=\sum_rX_{k}^{r\ast}\ A_{{\bf k}r}$, we get from Eq.(\ref{ExVib}) the polariton-vibration interaction by
\begin{eqnarray}
H_{pol-vib}&=&\sum_{\bf k,q}\sum_{r,s}\hbar\left\{F^{e}({\bf k+q})\ c_{\bf q}+F^{g}({\bf k})\ b_{\bf q}\right. \nonumber \\
&+&\left.F^{e}({\bf k})\ c_{\bf -q}^{\dagger}+F^{g}({\bf k+q})\ b_{\bf -q}^{\dagger}\right\} \nonumber \\
&\times&\left(X_{k}^{r\ast}X_{(k+q)}^{s}\right)\ A_{({\bf k+q})s}^{\dagger}A_{{\bf k}r},
\end{eqnarray}
where the exciton amplitudes indicate that the interaction is due to the polariton excitonic parts. The interaction describes scattering of polaritons between states with different wave vectors by the emission or absorption of vibrations, see figure (4). As the vibrations are dispersionless, one need to take care only for the energy conservation. The scattering take place between different momentum states in the lower and the upper polariton branches. But if the vibration energy equals the splitting energy between the two branches, polaritons can jump between the two branches by the emission or absorption of vibrations.

\begin{figure}
\centerline{\epsfxsize=8cm \epsfbox{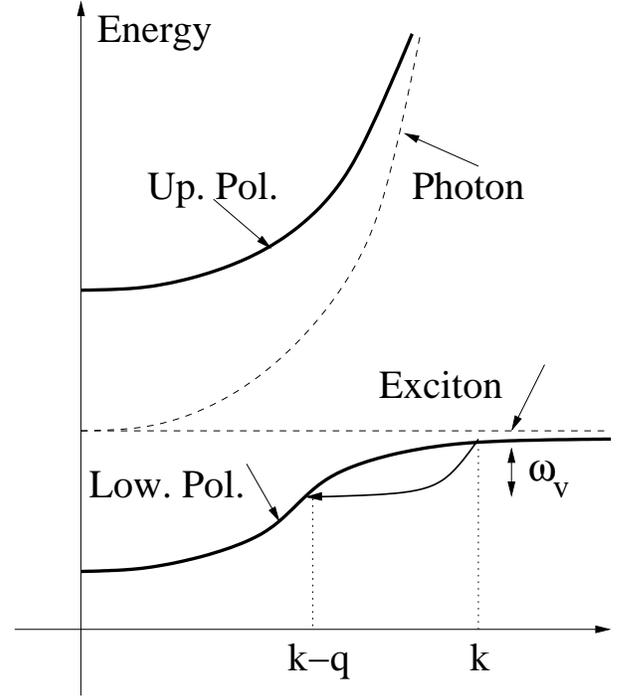}}
\caption{Schematic plot of the two polariton branches, with the cavity photon and exciton dispersions. The emission of a vibration by the scattering of a polariton between the wave vectors $k$ and $k-q$ is plotted.}
\end{figure}

The polariton-vibration interaction can serve as a significant source for polaritons relaxation toward the minimum energy at $k=0$. Using the Fermi golden rule, the damping rate for the spontaneous emission of vibrations at zero temperature $T=0$ off polaritons at the lower branch is given by 
\begin{equation}
w_{\bf k,q}^{--}\propto\left(\left|F^{e}({\bf k})\right|^2+\left|F^{g}({\bf k-q})\right|^2\right)\ \left|X_{k}^{-}\right|^2\left|X_{(k-q)}^{-}\right|^2.
\end{equation}
The process is plotted schematically in figure (4). The appearance of the exciton amplitudes indicate that the process is much more efficient at regions where the polaritons are more excitonic than photonic. The present relaxation mechanism does not suffer from the bottle-neck effect which appears in semiconductors due to the interaction of polaritons with acoustic phonons. Here the vibrations are dispersionless, while acoustic phonons have linear dispersion. The drawback of the present mechanism is the heating of the optical lattice atoms via exciting vibrational modes.

\section{Summary}

We analyzed a new mechanism for the coupling of vibrational modes in optical lattices to internal electronic excitations. Using perturbation theory, we derive the excitation-vibration interaction, which can be stronger than other known mechanisms via tunneling and scattering interactions. The excitation transfer among nearest neighbor sites is accompanied by the emission and absorption of a vibrational quanta. If the optical lattice is located within a cavity, in the strong coupling regime, the cavity photons and excitons form cavity polaritons, and where the exciton-vibration interactions result in polariton-vibration interactions. Such interaction will serve as an important source for relaxation of cavity polaritons towards their minimum energy at the lower polariton branch.


\ 

The work was supported by the Austrian Science Funds (FWF), via the Lise-Meitner Program (M977) and the project (P21101).

\end{document}